\documentclass[aps,pra,twocolumn,groupedaddress]{revtex4-2}

\usepackage{graphicx}
\usepackage{dcolumn}
\usepackage{bm}
\usepackage{amsmath}
\usepackage{lineno}
\begin{document}


\title{Boosting the photon bunching of classical light via a modified Fabry-P\'erot cavity}


\author{Lu Zhang$^{1,2}$}
\author{Dongxu Zhou$^1$}
\author{Hongzhi Zhang$^1$}
\author{Guoquan Zhang$^1$}
 \email[]{zhanggq@nankai.edu.cn}
\affiliation{$^1$The MOE Key Laboratory of Weak-Light Nonlinear Photonics, School of Physics and TEDA Applied Physics Institute, Nankai University, Tianjin 300071, China\\
$^2$School of Physics and Electronic Engineering, Jining University, Qufu 273155, China}


\date{\today}

\begin{abstract}
We proposed to boost the photon bunching effect of classical light via a modified Fabry-P\'erot cavity with one of the cavity mirrors being replaced by a spatial light modulator, through which multiple independent transverse modulations, each was to encode independently spatially correlated phase on the wave front of the light field, was cascaded in sequence in the longitudinal propagation direction in the cavity.
The two-photon bunching peak and the three-photon bunching peak of the generated superbunching light field were measured to be $36.9\pm2.0$ and $\rm (2.46\pm0.12)\times10^{3}$, respectively, which, as compared to those of thermal light, were improved by one-order and three-order of magnitude, respectively. Correspondingly, the ghost image visibility reconstructed  with the superbunching light was demonstrated experimentally to be improved by $\sim 40$ times as compared to that with the pseudo-thermal light. Such classical superbunching light field is of essential importance  not only for fundamental physics but also for potential applications such as correlated imaging, ghost interference and nonlinear optics.
\end{abstract}


\maketitle

\section{Introduction}
 Two-photon bunching effect was discovered by Hanbury Brown and Twiss (HBT)~\cite{brown1956correlation,brown1956test}, where a pair of photons from thermal light source had two alternative  indistinguishable two-photon paths to trigger a coincidence count of correlation measurement~\cite{Glauber1963PR,glauber2006nobel}. The HBT interferometer, as well as other similar configurations such as the correlated imaging schemes~\cite{Pittman1995PRA,strekalov1995PRL,Gatti2004PRL,Scarcelli2004EPL,Valencia2005Two,shapiro2008computational,bromberg2009ghost,kuplicki2016high,liu2018high} and the Hong-Ou-Mandel interferometer~\cite{Hong1987PRL}, are important two-photon interference devices. In principle, the two-photon interference is a result of superposition of the probability amplitudes of all possible indistinguishable two-photon paths according to Feynman's path integral theory~\cite{Liu2010PRA}, and it can be characterized by the normalized second-order correlation function $g^{(2)}(x_1,x_2)$ introduced by Glauber~\cite{Glauber1963PR}, in which $x_1$ and $x_2$ are the spatial positions of two photo-detectors $D_1$ and $D_2$. It is known that $g^{(2)}(x_1,x_2)$ is  always equal to 1 for coherent light, while there is a bunching peak of 2 at $x_1-x_2=0$ for the thermal light~\cite{Glauber1963PR,Scarcelli2004EPL}. This means that photons of thermal light have the tendency to come in bunches~\cite{Glauber1963PR,Scarcelli2004EPL}. In the case when $g^{(2)}(x_1-x_2=0)$ is larger than 2, the theoretical limit for thermal light, one says that the superbunching effect is achieved, which is of essential importance for applications such as ghost interference, ghost imaging~\cite{Pittman1995PRA,strekalov1995PRL,Gatti2004PRL,Scarcelli2004EPL,Valencia2005Two,shapiro2008computational,bromberg2009ghost,kuplicki2016high,liu2018high,zhou2024high} and multi-photon nonlinear light-matter interaction~\cite{Shen1967PR,Lambropoulos1968PR,Mollow1968PR,Agarwal1970PRA,Jechow2013NP}.

Over the past couple of decades, the superbunching effect was observed in various sources such as the collective two-level atoms coupled with a cavity~\cite{Dicke1954PR,Swain2000PRA,Temnov2009OE,Auffeves2011NJP,Hoi2012PRL,Grujic2013PRA,Bhatti2015SR}, the cavity-coupled quantum dot nanolasers~\cite{Leymann2015PRL,Jahnke2016NC,Redlich2016NJP},  the quantum light sources~\cite{Iskhakov2012OL,RN540} and even in some extreme phenomena like the optical rogue waves~\cite{arecchi2011granularity,leonetti2015observation}. In addition, light fields of the superbunching effect can be generated by modulating the amplitude and the phase of coherent light~\cite{Zhou2017PRA,Bai2017JOSAB,Strake2018OE,Liu2013AIP,bromberg2014generating,bender2018customizing,zhang2019superbunching,RN306,liu2021simple}.
In general, from the viewpoint of quantum optical coherence, the key to achieve the two-photon superbunching effect is to introduce more than two indistinguishable two-photon paths triggering a two-photon coincidence count~\cite{Hong2012PRA,hong2013high,zhang2019superbunching,RN306,luo2021two,heimerl2024multiphoton,heimerl2025quantum,PhysRevLett.119.223603,manceau2019indefinite}. This can be achieved by modulating the light field in the longitudinal dimension, as demonstrated by inserting optical nonlinear medium~\cite{Bromberg2010NP}, optical channels of first-order incoherence~\cite{Hong2012PRA}, rotating ground glasses~\cite{Zhou2017PRA,Bai2017JOSAB}, and acousto-optic modulators~\cite{Strake2018OE} in the propagation optical path. It can also be realized by modulating the light fields in the transverse dimension, for instance, by modulating the wave front of the light field with spatial light modulator (SLM)~\cite{bromberg2014generating,bender2018customizing,zhang2019superbunching}.
One may note that it is relatively easy to obtain a light field with a high bunching peak value in the temporal domain, for example, the temporal two-photon bunching peak $g^{(2)}(t=0)$ can reach $\sim 350$ by using an acousto-optical modulator to increase the temporal intensity fluctuation of light field~\cite{Strake2018OE}. However, this is not the case in the spatial domain, where the bunching peak value of the normalized second-order correlation function is relatively low, no matter in the linear or nonlinear systems~\cite{Liu2013AIP,Hong2012PRA,Bromberg2010NP}. Nevertheless,  light fields with a two-photon bunching peak higher than 10  were reported in the spatial domain with the digitally designed spatially phase-correlated wave front~\cite{zhang2019superbunching}. Note that light fields with a higher two-photon or multi-photon superbunching peak are highly desired for improvement on the performance of practical applications such as ghost imaging, ghost interference and higher-order nonlinear light-matter interaction.

In this article, we proposed a method to boost the bunching effect of classical light via a modified Fabry-P\'{e}rot (F-P) cavity, through which light fields can be modulated in both transverse and longitudinal dimensions, therefore, to introduce multiple different but indistinguishable two-photon paths. In addition, we found that not only the two-photon bunching peak but also the multi-photon bunching peak were increased significantly as compared to those of thermal light. Ghost imaging with high visibility was also demonstrated experimentally.  The ability to generate classical light field with superbunched photons is of both fundamental importance and potential application prospects.

\section{Theoretical Model}
\begin{figure}[t]
\centering\includegraphics[width=6cm]{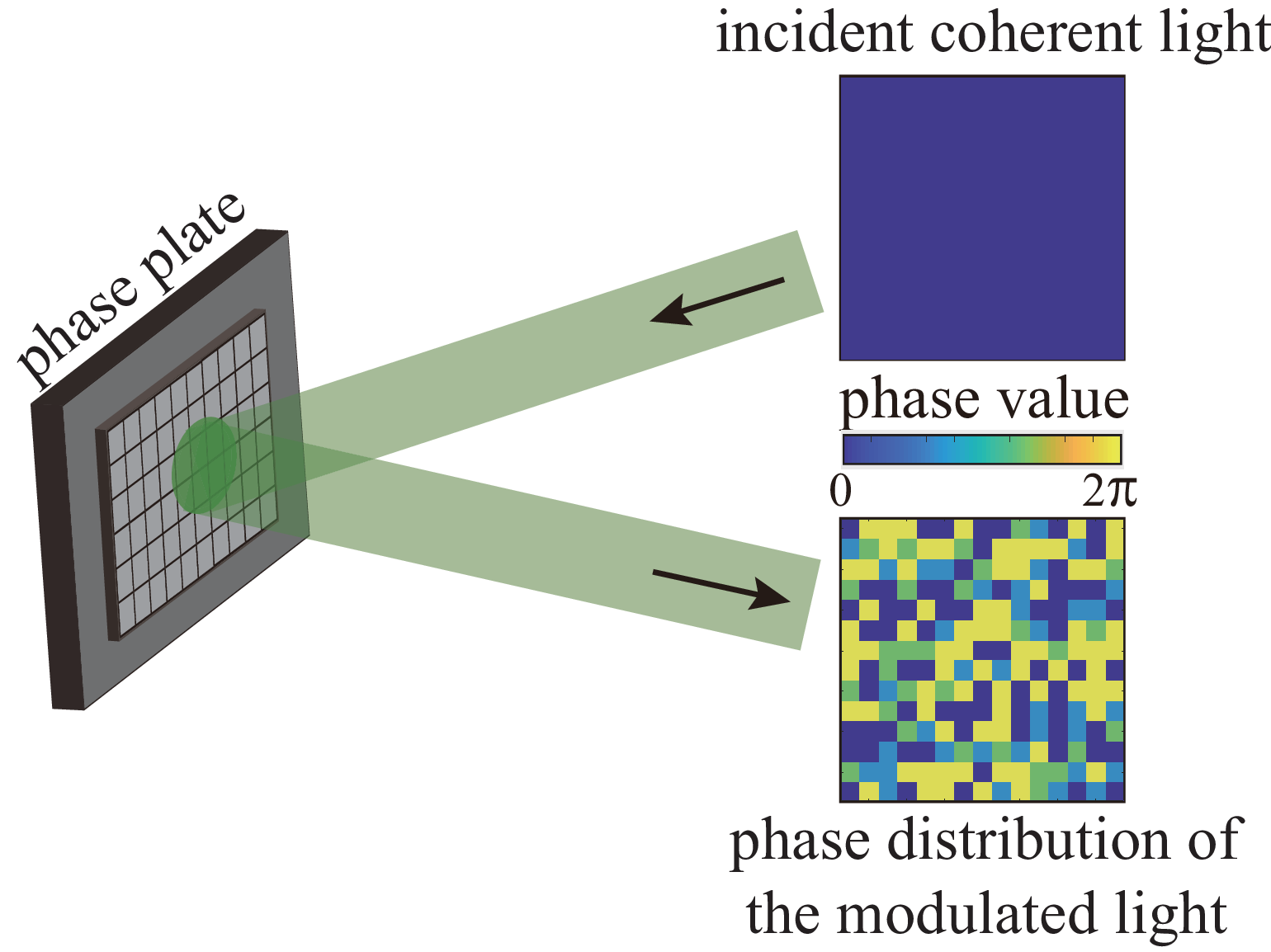}
\caption{A schematic diagram to encode the spatially correlated phase structure on the wave front of the light field.}\label{fig:slm_modulation}
\end{figure}

According to Feynman's path integral theory~\cite{Liu2010PRA}, there are two different but indistinguishable two-photon paths  $AD_1BD_2$ and $AD_2BD_1$  in the traditional bunching effect of thermal light~\cite{brown1956correlation,brown1956test}, where $AD_1BD_2$ represents the case that the photon $A$ from the source plane is detected by the detector $D_1$ in the detection plane while the photon $B$ is detected by the detector $D_2$, and $AD_2BD_1$ depicts the similar case that the photon $A$ is detected by the detector $D_2$ and the photon $B$ is detected by the detector $D_1$. The superposition of these two alternative indistinguishable two-photon paths will result in a bunching effect with a bunching peak of 2 at $x_1=x_2$, showing the tendency of radiated photons of thermal light to come in bunches rather than randomly, as is the case for the coherent light~\cite{brown1956correlation,brown1956test,Glauber1963PR}. It has been demonstrated that the two-photon bunching effect can be improved by introducing additional indistinguishable two-photon paths~\cite{Hong2012PRA,hong2013high,zhang2019superbunching,RN306,luo2021two}.

Previously, we designed a classical light field with spatially phase-correlated wave front by using a digital wave front encoding technique, demonstrating a spatial superbunching effect~\cite{zhang2019superbunching}. The schematic configuration is shown in Fig.~\ref{fig:slm_modulation}, in which a phase plate (a SLM in experiments) is used to encode the specially designed phase distribution  dispersed randomly with $N$ spatially correlated phases $\phi$, $2\phi$, $3\phi$, $\cdots$, $N\phi$ onto the wave front of a coherent light, where $\phi$ is a temporally random phase with equal distribution probability within $[0,2\pi)$~\cite{zhang2019superbunching}. In this way, multiple different but indistinguishable two-photon paths can be introduced whenever the amplitudes of the two-photon paths  $A_mD_1A_nD_2$ and $A_mD_2A_nD_1$ are of the same encoded random phase factor $(m+n)\phi=l\phi$ ($m, n=1,\,\cdots,\, N$), where $m\phi$ and $n\phi$ are  the encoded random phase of photons $A_m$ and $A_n$, respectively. The superposition of all these different but indistinguishable two-photon paths with the same random phase factor $l\phi$ and then summation over all probability contributions from all possible cases with different $l\phi$ will result in a superbunching effect. The normalized second-order spatial correlation function is~\cite{zhang2019superbunching}

\begin{widetext}
\begin{eqnarray}\label{Eq:coincidentrateWAVEFRONT}
g^{(2)}_N(x_1, x_2)&&=\frac{1}{N^2}\sum_{l'=-N+2}^{N+1}\left\{(l'-1)+2\left\lfloor\frac{l'-1}{2}\right\rfloor\text{sinc}^2\left[\frac{kR}{2z}(x_1-x_2)\right]\right.\nonumber\\
&&\left.+\left[(l'-1)(l'-2)-2\left\lfloor\frac{l'-1}{2}\right\rfloor\right]\text{sinc}^2\left(\frac{kx_1R}{2z}\right)\text{sinc}^2\left(\frac{kx_2R}{2z}\right)\right\},
\end{eqnarray}
\end{widetext}

\noindent
where $\left\lfloor\ \cdots\right\rfloor$ means to round the internal numbers downward, $R$ is the cross-section diameter of the light source,  $z$ is the longitudinal distance between the light source and the detection plane, and $k$ is the wavenumber of the light. Theoretically, the normalized correlation curve is bunched at $x_1=x_2=0$ with a peak value of $g^{(2)}_{N}(x_1= x_2=0)=\left(\frac{2N^2+1}{3N}\right)$.
With this scheme, the superbunching effect with a superbunching peak of $11.72\pm0.05$ was obtained in experiments~\cite{zhang2019superbunching}.

\begin{figure}[b]
\centering\includegraphics[width=8cm]{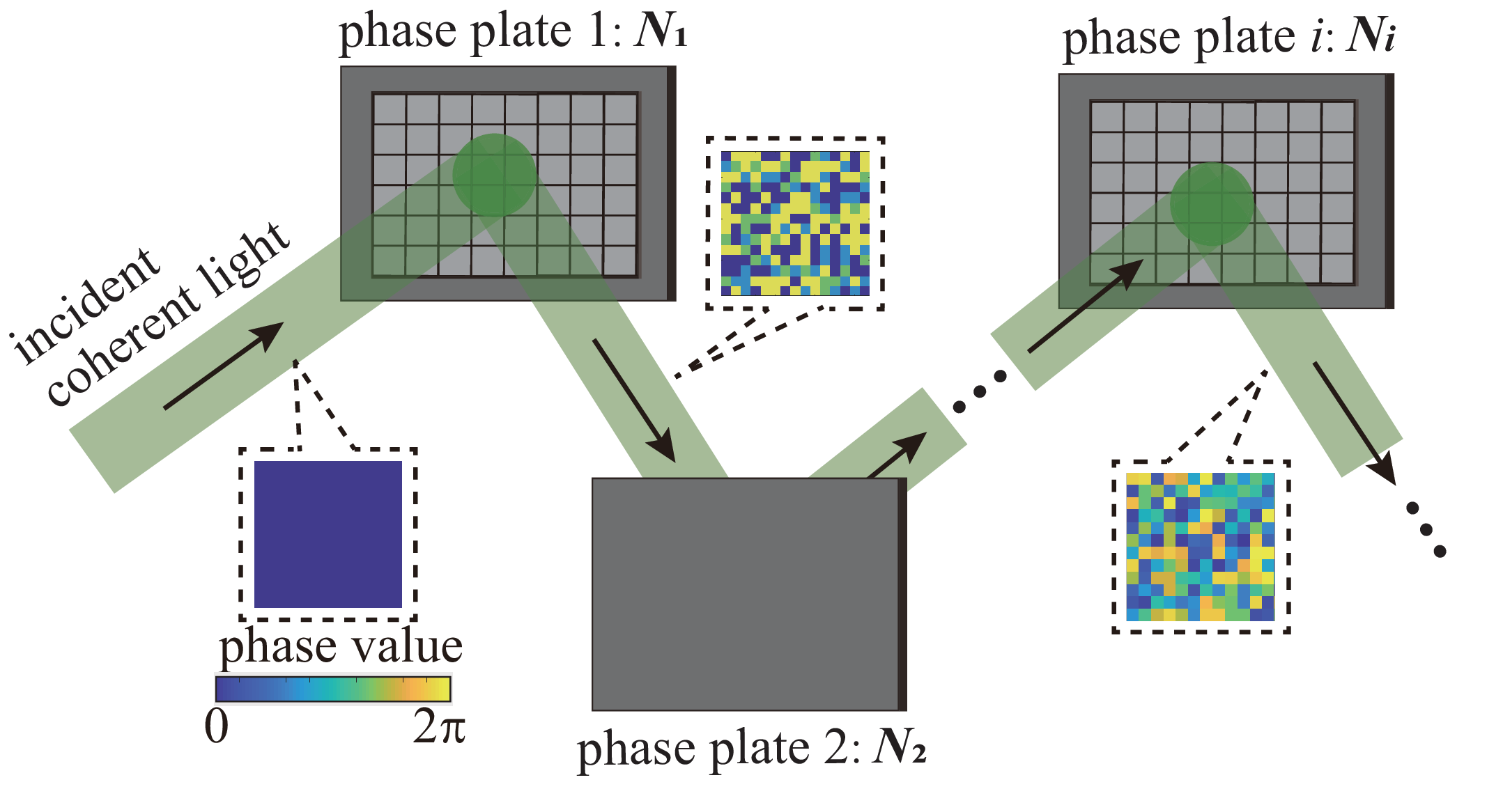}
\caption{Schematic diagram to modulate the wave front of the light field through  a series of independent phase plates cascading in the longitudinal propagation direction. The $i$th phase plate will encode $N_i$ spatially correlated phases on the wave front of the light fields.}\label{fig:multi_modulation}
\end{figure}

Now let one cascade the phase plates in the longitudinal propagation dimension, as shown in Fig.~\ref{fig:multi_modulation}, in which the $i$th phase plate will encode $N_i$ spatially correlated phases on the wave front of the incident light field. It is clear that the two-photon paths $A_{m,\cdots,p,\cdots}D_1A_{n,\cdots,q,\cdots}D_2$ satisfying $m\phi+n\phi+\cdots+p\phi'+q\phi'+\cdots=\mathrm{constant}$ are indistinguishable, where $\phi$ and $\phi'$ are independent temporal random phases distributed equally within $[0,2\pi)$, respectively. Here, the two-photon path $A_{m,\cdots,p,\cdots}D_1A_{n,\cdots,q,\cdots}D_2$ indicates the case that the photon $A_{m,\cdots,p,\cdots}$ encoded with a random phase  $m\phi+\cdots+p\phi'+\cdots$ by the phase plates 1 ($m\phi$), $\cdots$, $i$ ($p\phi'$), $\cdots$ in sequence is detected by detector $D_1$ and the photon $A_{n,\cdots,q,\cdots}$ encoded with a random phase  $n\phi+\cdots+q\phi'+\cdots$ by the  phase plates 1 ($n\phi$), $\cdots$, $i$ ($q\phi'$), $\cdots$ in sequence is detected by detector $D_2$.  Similarly, the two-photon correlation count rate can be calculated by superposing all these different but indistinguishable two-photon paths with the same random phase factor  and then summing over all probability contributions from all possible cases with different random phase factor. For simplicity but without loss of generality, suppose that there are $M$ phase plates, each of them is of independent $N$ spatially correlated phase structure. In this case, after tedious but straightforward calculation (see Appendix for detailed rigorous derivation), the normalized second-order spatial correlation function can be expressed as

\begin{widetext}
\begin{eqnarray}\label{Eq:coincidentrate-M}
g^{(2)}_{N,M}(x_1, x_2)&&=\frac{1}{N^{2M}}\prod^{M}_{i=1}\sum_{l_i=-N+2}^{N+1}\left\{(l_i-1)+2\left\lfloor\frac{l_i-1}{2}\right\rfloor \mathrm{sinc}^2\left[\frac{kR}{2z}(x_1-x_2)\right]\right.\nonumber\\
&&\left.+\left[(l_i-1)(l_i-2)-2\left\lfloor\frac{l_i-1}{2}\right\rfloor\right]\mathrm{sinc}^2\left(\frac{kx_1R}{2z}\right)\mathrm{sinc}^2\left(\frac{kx_2R}{2z}\right)\right\}.
\end{eqnarray}
\end{widetext}

In comparison with Eq.~(\ref{Eq:coincidentrateWAVEFRONT}), one sees that the normalized second-order  spatial correlation function $g^{(2)}_{N,M}(x_1, x_2)$ is simply a  product of $M$ cascading $g^{(2)}_{N_i}(x_1, x_2)$ with $N_i$ being the number of spatially correlated phases on the $i$th phase plate (here $N_i=N$),  similar to the case in our previous work by inserting $M$ pairs of mutually first-order incoherent optical channels in sequence into the traditional HBT interferometer~\cite{Hong2012PRA}. Intuitively, for the two-photon bunching effect with a cascading configuration of multiple phase plates with spatially correlated phases, each pair of photons incident on a phase plate will obey exactly the same bunching enhancement mechanism and experience exactly the same bunching enhancement effect, therefore, the cascading configuration of multiple phase plates with spatially correlated phases will result in a cumulative enhancement on the photon bunching effect. The bunching peak of the normalized second-order spatial correlation function at $x_1=x_2=0$ can be obtained

\begin{equation}\label{Eq:peakT}
g^{(2)}_{N,M}(x_1=x_2=0)=\left(\frac{2N^2+1}{3N}\right)^{M}.
\end{equation}

In the following section, we will give an experimental verification on the proposed spatial two-photon superbunching effect generated by cascading multiple independent phase plates with spatially correlated phase structure. In addition, the three-photon spatial superbunching effect will also be demonstrated experimentally.

\section{Experimental verification and discussions}\label{Sec:experiment}

\begin{figure}[b]
\centering\includegraphics[width=8cm]{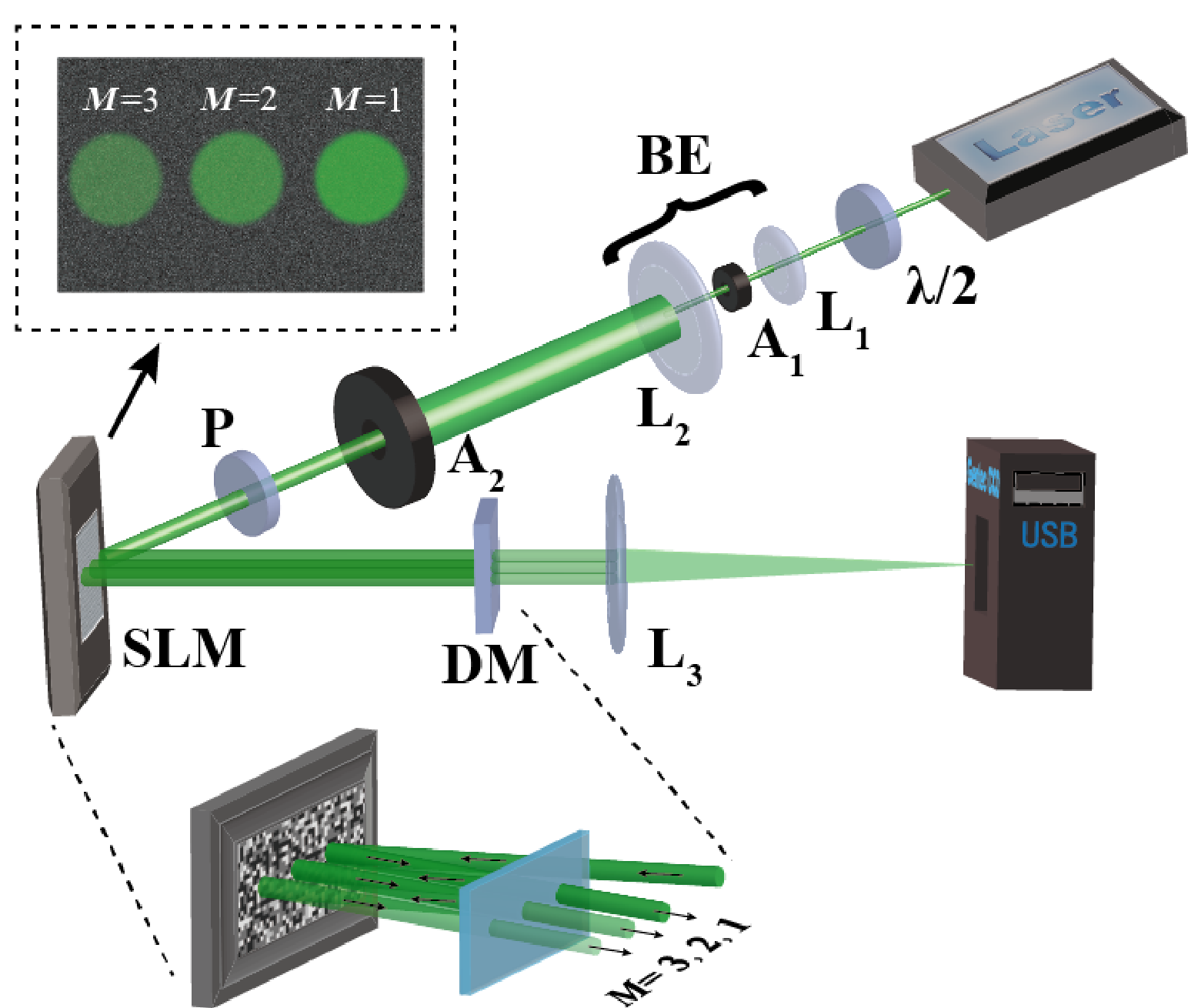}
\caption{Schematic diagram of the experimental setup. $\lambda/2$: half-wave plate; L$_1$, L$_2$ and L$_3$: lens with focal length 10 mm, 120 mm and 800 mm, respectively; A$_1$ and A$_2$: iris; BE: beam expender; P: polarizer; SLM: spatial light modulator; DM: dielectric mirror; CCD: charge coupled device. The inset on the up-left corner shows the light spots on the CCD. The bottom inset shows the detail of the cascading phase plates in the modified F-P cavity.}\label{fig:scheme}
\end{figure}

Figure~\ref{fig:scheme} shows the schematic diagram of the experimental setup to modulate the wave front of the light field by a series of independent phase plates cascading in sequence in the longitudinal propagation dimension. Here, a single-mode continuous-wave laser light operating at 532 nm was expanded and collimated by a pair of lens L$_1$ and L$_2$ with focal lengths of 10 mm and 120 mm, respectively. The iris A$_1$ was used as a spatial filter to improve the beam quality. The light transmitted through another iris A$_2$ so that the diameter of the transmitted beam spot was set to be $R=5$ mm. The polarization and the intensity of the light was controlled by a combination of a half-wave plate  $\lambda/2$ and a polarizer P, and the light was set to be horizontally polarized. The light was coupled obliquely into a modified F-P cavity, which was composed of a reflection-type phase-only SLM (HEO 1080P from HOLOEYE Photonics AG, Germany)  and a dielectric mirror (DM). The reflectivity of the SLM at 532 nm was about 65\%, and that of the DM was  99\%, so the finesse of the F-P cavity was only about  14.2. Note that the light was incident obliquely onto the SLM and the DM so that the light was reflected multiple times within the F-P cavity while the beam spots on the SLM and the DM were separated in space with respect to each other, as shown in the insets of Fig.~\ref{fig:scheme}. The spatially correlated phases were encoded onto the wave front of the light fields through the SLM whenever the light was reflected from the SLM, just as the case in our previous work~\cite{zhang2019superbunching}. Note that each element of SLM could be controlled independently, therefore, the spatially correlated phases encoded on the wave front of the light reflected from the SLM could also be independently designed. The light beams output from the modified F-P cavity were collected by lens L$_3$ with a focal length of 800 mm and detected by a CCD camera placed on the focal plane of lens L$_3$. In this way, one can cascade the phase plates in sequence in the longitudinal propagation direction via the F-P cavity, and the number of the phase plates $M$ is determined by the reflection times from the SLM in the F-P cavity, as shown in the insets of Fig.~\ref{fig:scheme}. Here, we should emphasize that the modified F-P cavity was used as a convenient and effective way to cascade multiple phase plates with spatially correlated phases, which is totally different from the role of the traditional F-P interferometer in the scope of the first-order coherence.

The high-order coherence of the light fields output from the modified F-P cavity was measured. Here, by using the digital wave front encoding technique developed in our previous work~\cite{zhang2019superbunching}, one modulated the wave front of the light fields using the SLM with specially designed phase distribution dispersed randomly with $N$ spatially correlated phases $\phi_i$, $2\phi_i$, $3\phi_i$, $\cdots$, $N\phi_i$, where $\phi_i$ ($i=1, \, 2,\, \cdots,\, M$) was a temporally random phase with equal distribution probability within [0, $2\pi$). The intensity pattern of the $i$th beam spot was measured using the CCD camera, each frame, with an acquisition time of 1 ms, corresponding to the case with a random phase $\phi_i$. The normalized second-order spatial correlation function can be calculated  through the formula $g^{(2)}_{N,M}(x_1,x_2)=\left<I_{N,M}(x_1)I_{N,M}(x_2) \right>/\left< I_{N,M}(x_1)\right>\left< I_{N,M}(x_2)\right>$
based on a collection of 10,000 frames, where $I_{N,M}(x)$ is the intensity at position $x$, and $\left<\cdots\right>$  represents the ensemble average.

\begin{figure}[htbp]
\centering\includegraphics[width=9cm]{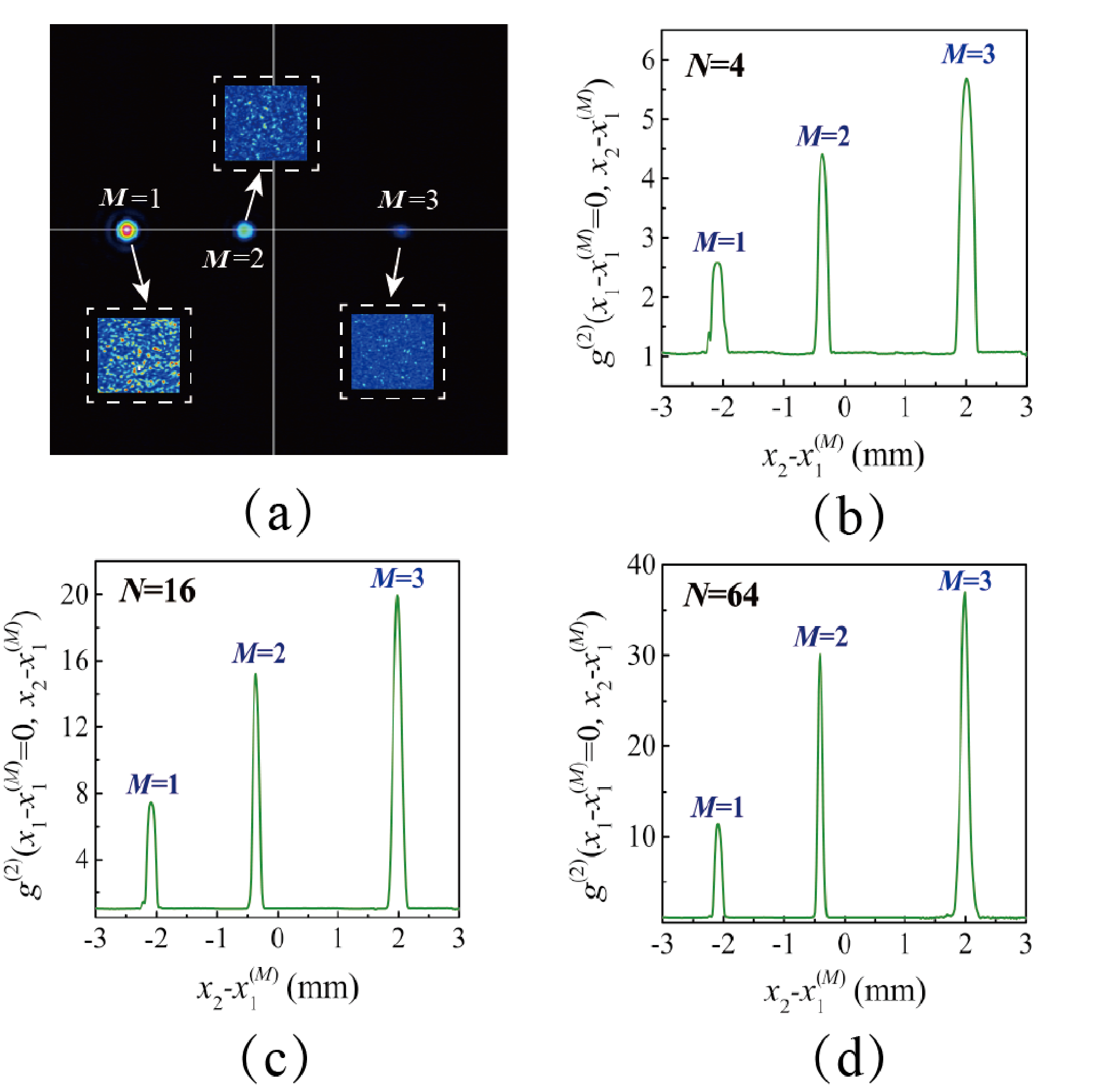}
\caption{(a) The beam spots and the intensity distribution within the beam spots recorded by the CCD camera. (b), (c) and (d) show the normalized second-order spatial correlation functions of the light fields ($M=1,\,2,\,3$) output from the modified F-P cavity with $N=4,\,16,\, 64$, respectively. Note that the transverse coordinates are spatially shifted by $x_1^{(M)}$.}\label{fig:M123}
\end{figure}

\begin{figure}[htbp]
\centering\includegraphics[width=6.5cm]{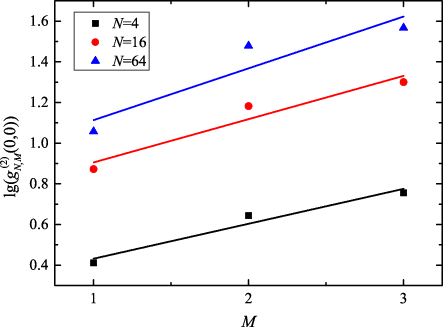}
\caption{The measured bunching peak values  of the light fields output from the modified F-P cavity with different $N$ and $M$, in which the black squares, the red dots and the blue triangles are for the cases with $N=4$,~16 and 64, respectively. The lines are linear fits to the measured data according to Eq.~(\ref{Eq:peakT}). }\label{fig:comparison}
\end{figure}

\begin{figure}[htbp]
\centering\includegraphics[width=6.5cm]{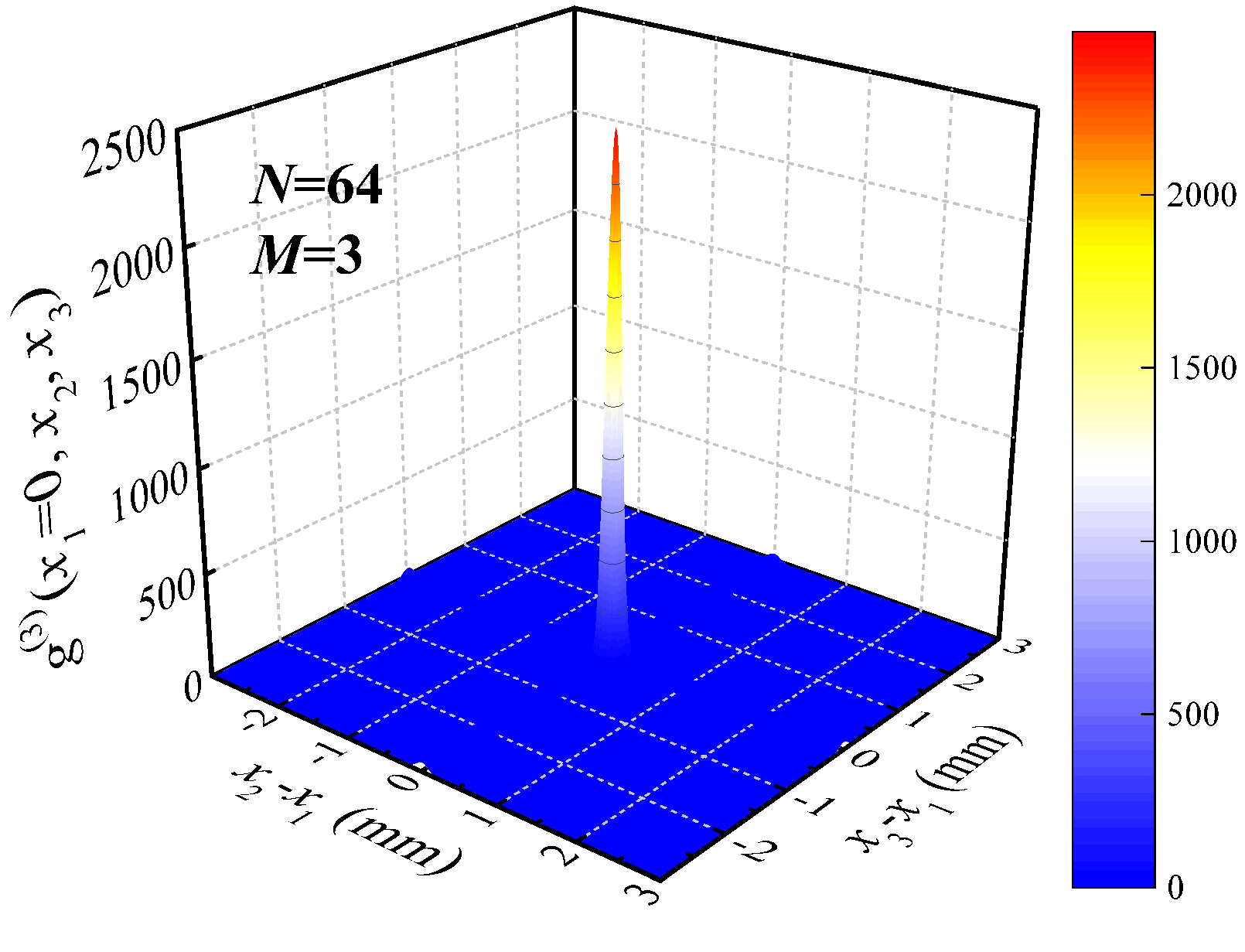}
\caption{The measured normalized third-order spatial correlation function of the light field output from the modified F-P cavity in Fig.~\ref{fig:scheme} with $N=64$ and $M=3$. }\label{fig:g3}
\end{figure}

\begin{figure*}[htb]
\centering\includegraphics[width=17cm]{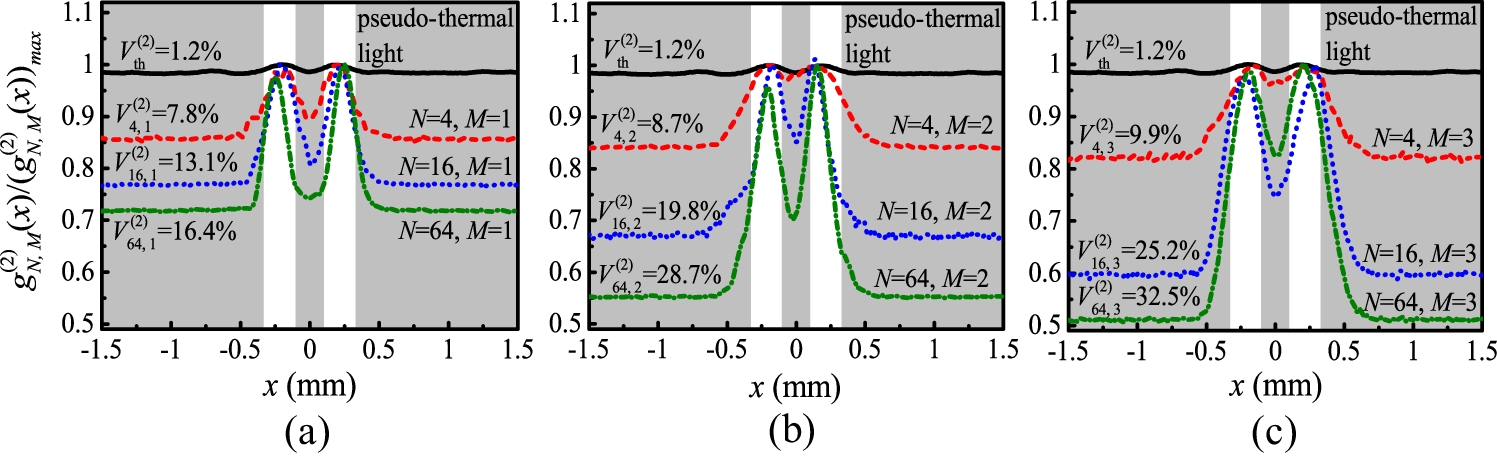}
\caption{(a)-(c) The normalized cross-sectional profiles of the second-order ghost images employing the pseudo-thermal field and the superbunching fields with  $M=1$, 2 and 3, respectively. The black solid curves are for the cases with the pseudo-thermal light field, and the red dashed curves, the blue dotted curves and the green dot-dashed curves are for the cases with the superbunching fields with $N=4$, 16 and 64, respectively. }\label{fig:2orderGI}
\end{figure*}

Figure~\ref{fig:M123} shows the measured intensity distribution of the beam spots on the CCD camera and the normalized second-order spatial correlation function with different cascading number $M$ of the phase plates. Note that the number of phase plates $M$ in Eq.~(\ref{Eq:coincidentrate-M}) is indeed the times of reflection from the SLM in experiments, so the first beam output from the F-P cavity corresponds to the case with $M=1$, and the second beam output from the F-P cavity refers to the case with $M=2$, etc., as shown in Fig.~\ref{fig:M123}(a), where totally three beam spots are shown because of the space limitation of SLM and DM. It is evident that the output light intensity decreases with the increase of $M$ due to the reflection loss of both SLM and DM. The intensity pattern within each beam spot is shown in the inset of Fig.~\ref{fig:M123}(a). The measurement results on the normalized second-order spatial correlation functions with different   $N$ and  $M$ are shown in Figs.~\ref{fig:M123}(b)-\ref{fig:M123}(d), respectively. Note that each bunching curve is shifted in space by $x_1^{(M)}$ in the transverse dimension. It is seen that  when $N=4$, the superbunching peaks of $g^{(2)}_{4,M}(x_1-x_1^{(M)}=0, x_2-x_1^{(M)}=0)$ for the cases with $M=1, \,2,\, 3$ were measured to be $2.58\pm0.13$, $4.40\pm0.22$ and $5.69\pm0.28$, respectively. When $N=16$, the superbunching peaks of $g^{(2)}_{16,M}(x_1-x_1^{(M)}=0, x_2-x_1^{(M)}=0)$ for the cases with $M=1, \,2,\, 3$ were measured to be $7.47\pm0.37$, $15.18\pm0.76$ and $19.9\pm1.0$, respectively. And when $N=64$, the superbunching peaks of $g^{(2)}_{64,M}(x_1-x_1^{(M)}=0, x_2-x_1^{(M)}=0)$ for the cases with $M=1, \,2,\, 3$ were measured to be $11.42\pm0.57$, $30.1\pm1.5$ and $36.9\pm2.0$, respectively. One sees that the superbunching effect is achieved in each case,  the bunching peak value increases with increasing $N$ and $M$,  and the logarithmic values of the bunching peaks increase linearly with $M$,  in good agreement with the theoretical prediction by Eq.~(\ref{Eq:peakT}),  as shown in Fig.~\ref{fig:comparison}. For comparison, we also measured the normalized second-order spatial correlation function of pseudo-thermal light produced by encoding spatially distributed random phase on the wave front of light with $M=1$, and the bunching peak was measured to be $1.69\pm0.10$. Interestingly, we confirmed experimentally that the bunching peak of pseudo-thermal light will not change with the increase of $M$. This is because the phase encoded on the wave front of light is random without any correlation in spatial domain, and cascading the phase plates, each with a random phase distribution, will not change the randomness property of the phases carried by the light.

In principle, similar to the case of the second-order coherence, multiple indistinguishable three-photon paths can also be introduced and its number increases significantly by encoding the wave front of light with spatially correlated phases in the transverse dimension and then cascading the effect in the longitudinal propagation dimension through the modified F-P cavity. This was confirmed experimentally by measuring the third-order spatial correlation function of the light fields  output from the modified F-P cavity shown in Fig.~\ref{fig:M123}. As a typical example, Fig.~\ref{fig:g3} shows the measured normalized third-order spatial correlation function with $N=64$ and $M=3$.  One sees that the three-photon bunching peak $g^{(3)}_{64,3}(x_1= x_2= x_3=0)$ was measured to be as high as  $\rm (2.46\pm0.12)\times 10^{3}$, much higher than the three-photon bunching peak of thermal light, which is theoretically limited to 6 and was measured experimentally to be $3.46\pm0.17$ when a spatially random phase was encoded on the light field with $M=1$.

\section{Applications on ghost imaging}\label{Sec:Applications}

The visibility is an important parameter to evaluate the quality of an image, this is also true for ghost images. The visibility of ghost images is determined by the contrast of the normalized high-order correlation function of the light field, and a higher bunching peak value usually means a higher contrast of the normalized high-order correlation function, therefore, a higher visibility of ghost images.
The bunching peak value of the normalized second-order correlation function for the bunched non-classical light such as entangled photon pairs is much higher than that of classical light such as thermal light, and theoretically the visibility of the ghost images using bunched non-classical light can reach 100\%~\cite{strekalov1995PRL}. In contrast, the visibility of ghost images reconstructed with thermal light is in general less than 33.3\%. For high-order coherence, the bunching peak value with bunched non-classical light is also much higher than that with thermal light, therefore, the visibility of ghost images reconstructed with bunched non-classical light is much higher than that reconstructed with thermal light.

Here, as a proof of principle, we demonstrated the ghost imaging with a significantly improved visibility when the superbunching fields were employed.
The object was a double-slit with a slit-width of 200 $\mu$m and a slit-separation-distance of 400 $\mu$m, respectively, as shown by the white stripes in Fig.~\ref{fig:2orderGI}. For comparison, both the pseudo-thermal light and the superbunching light produced as described above were used in the experiments. The typical conventional ghost imaging scheme~\cite{Pittman1995PRA,strekalov1995PRL} was adopted, in which the object beam transmitting through the double-slit was collected by a bucket detector without position information and the reference beam, which did not interact with the object, was detected by a detector with position information. The ghost images were reconstructed by spatially correlating the intensity distributions detected by the two detectors.
Note that, to suppress the scattering noise reflected from the SLM, we superposed a blazed grating phase onto the SLM to remove the stray light reflected from the SLM. In this way,  the visibility of ghost images was optimized even for pseudo-thermal light.

\begin{figure}[t]
\centering\includegraphics[width=6.5cm]{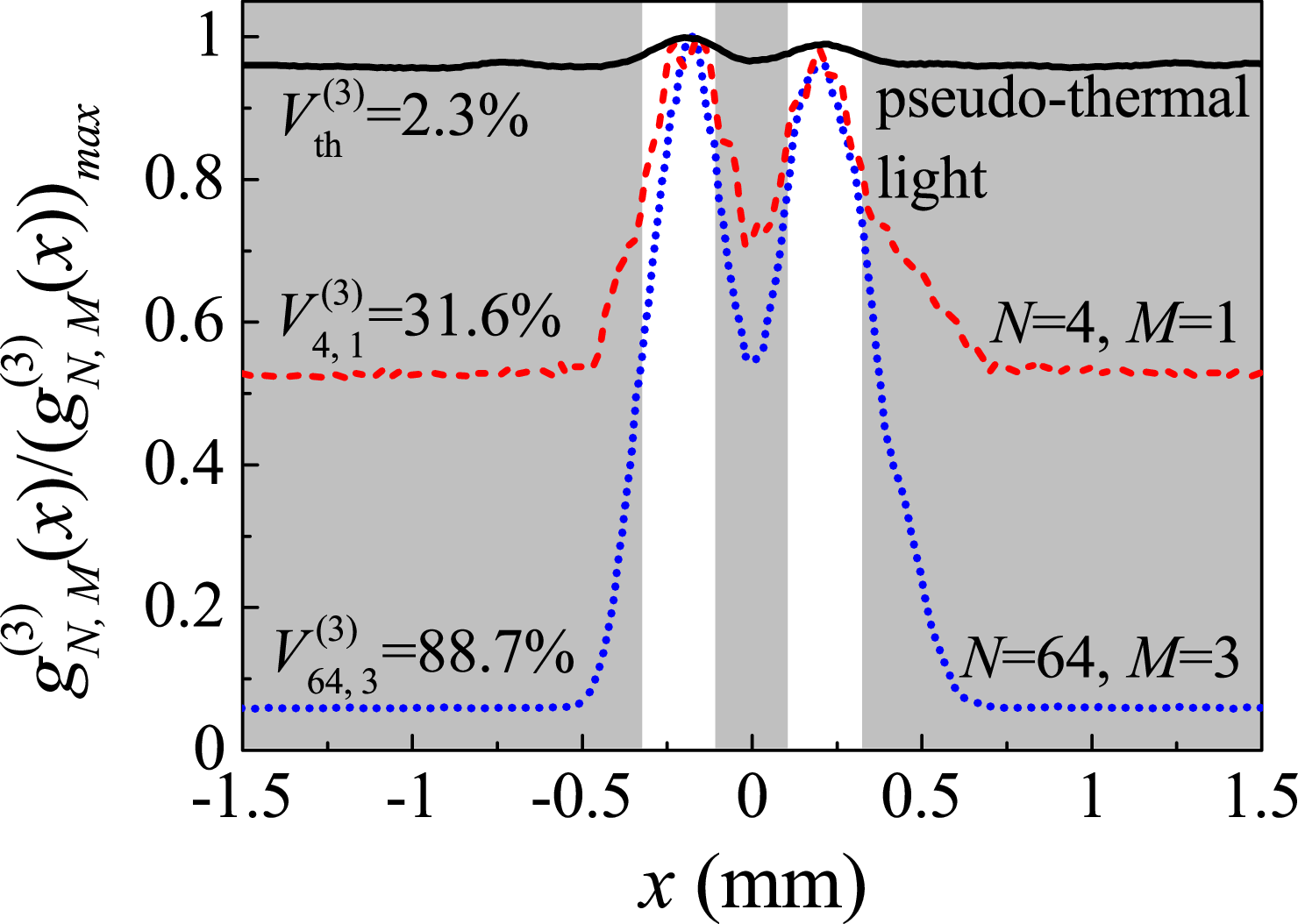}
\caption{The measured normalized cross-sectional profiles of the third-order ghost images with the pseudo-thermal light field and the superbunching fields with different $N$ and $M$. The black solid curve is for the case with the pseudo-thermal field, and the red dashed and the blue dotted curves are for the cases with the superbunching fields with $N=4$, $M=1$ and $N=64$, $M=3$, respectively.}\label{fig:3orderGI}
\end{figure}

Figure~\ref{fig:2orderGI} shows  the normalized cross-sectional profiles of the reconstructed second-order ghost images with the pseudo-thermal light and the superbunching light of different $N$ and $M$, respectively. As expected, one sees that the image visibility $V^{(2)}_{N,M}$ with the superbunching light  is significantly increased as compared to that with the pseudo-thermal light $V^{(2)}_{th}$, from $V^{(2)}_{th}=1.2$\% for the pseudo-thermal light to $V^{(2)}_{64,3}=32.5$\% for the superbunching light with $N=64$~and $M=3$. In addition, the image visibility for the superbunching light also increases with increasing $N$ and $M$, in good agreement with the predicted bunching property of the superbunching fields. Note that the image visibility for the pseudo-thermal light $V^{(2)}_{th}$ does not change with the parameters $N$ and $M$.

Moreover, we also measured experimentally the third-order ghost images with the pseudo-thermal light and the superbunching light, respectively, and the results are shown in Fig.~\ref{fig:3orderGI}. The image visibility of the third-order ghost image for the pseudo-thermal light $V^{(3)}_{th}$ was measured to be 2.3\%, which is increased as compared to $V^{(2)}_{th}=1.2$\% of the second-order ghost image with the same pseudo-thermal light. It is delighted to see that the visibility of the third-order ghost image $V^{(3)}_{N,M}$ was boosted to be 88.7\% by employing the superbunching field with $N=64$ and $M=3$.  We also confirmed experimentally that the visibility of the third-order ghost image for the superbunching light  $V^{(3)}_{N,M}$ increases with the increase of the parameters $N$ and $M$ of the superbunching light fields, as shown in Fig.~\ref{fig:3orderGI}, and again the visibility of the third-order ghost image for the pseudo-thermal light $V^{(3)}_{th}$ does not change with the parameters $N$ and $M$.

\section{Conclusion}

In conclusion, we boosted the multi-photon bunching effect of classical light through a modified F-P cavity, in which one of the cavity mirrors was replaced by a SLM so that the wave front of the light fields could be modulated to be partially spatially correlated in the transverse dimension via the SLM, and then the transverse wavefront modulations were cascaded in sequence in the longitudinal propagation direction. Classical superbunching light fields with a two-photon bunching peak of $36.9\pm2.0$ and a three-photon bunching peak of  $(2.46\pm0.12)\times10^{3}$, much higher than those of thermal light, were achieved experimentally, which theoretically can be improved  further on by cascading more phase plates with spatially correlated phases. By using the superbunching light fields, the visibility of the second- and third-order ghost images was confirmed experimentally to be improved by  $\sim 40$ times as compared to that using pseudo-thermal light fields.  Our method provides an effective way to bunch photons of classical light and could have potential applications such as correlated imaging, ghost interference and nonlinear optics.

\appendix*
\section{}
We will give the derivation of the two-photon probability distribution and the bunching effect of light fields based on Feynman’s path integral theory~\cite{Liu2010PRA}. Here, we only consider the problem in the spatial domain, and the detection time difference $\tau$ between two detectors for an effective coincidence count is set to be zero for simultaneous detection of photons. In the one-dimensional case and under the paraxial approximation,
the normalized second-order spatial correlation function of thermal light is

\begin{eqnarray}\label{Eq:A1}
g^{(2)}(x_1,x_2)&&=\frac{\sum\limits_{A,B\in R}P^{(2)}_{AB}(x_1,x_2)}{P^{(1)}(x_1)P^{(1)}(x_2)}\nonumber\\
&&=1+\mathrm{sinc}^2\left(\frac{k(x_1-x_2)R}{2z}\right).
\end{eqnarray}
\noindent
Here, the light field is assumed to be of single frequency, $k$ is the wave vector of the light propagating from the source plane to the detection plane, $R$ is the cross-section diameter of the light source, and $z$ represents the longitudinal distance between the light source and the detection planes. $P^{(1)}(x_q)=\left<\left|\sum_{X'\in R}e^{i\phi_{X'}}g_{X'q}\right|^2\right>=1/z^2$ is the photon counting rate of the detector $D_q$ ($q\in \{1,2\}$). $P^{(2)}_{AB}(x_1,x_2)=\left<\left|\sum_{j=1}^2 F(A,B)_j\right|^2\right>=\left|g_{A1}g_{B2}+g_{A2}g_{B1}\right|^2$ is the coincidence counting rate caused by the superposition of two different but indistinguishable two-photon paths. $F(A,B)_1=e^{i\phi_A}g_{A1}e^{i\phi_B}g_{B2}$ and $F(A,B)_2=e^{i\phi_A}g_{A2}e^{i\phi_B}g_{B1}$ are the probability amplitudes of the two-photon paths. $\left<\cdots\right>$ represents the ensemble average. $\phi_{X'}$ is the initial phase of the photon $X'$ ($X'\in \{A,B\}$), and $g_{X'q}=\frac{e^{-ik\left|r_q-r_{X'}\right|}}{\left|r_q-r_{X'}\right|}e^{i\omega t_q}$ is the propagation factor for photon $X'$ propagating from the source to the detector $D_q$. $\omega$ is the frequency of the light. $r_q$ is the coordinate of the detector $D_q$ and $r_{X'}$ is the coordinate of photon triggering a coincidence count. It’s worth noting that the spatial longitudinal propagation distances of the two photons triggering the two-photon coincidence count are equal, and the temporal terms  $e^{i\omega t_q}$ in their propagation factors are also the same due to the paraxial approximation and the single frequency assumption. As a result, the two-photon coincidence counting rate is independent of time as the calculation requires taking the modulus of the probability amplitude of the two-photon paths. Therefore, the normalized second-order correlation function is also independent of time, and Eq.~(\ref{Eq:A1}) results in a bunching curve with a bunching peak of 2, which is typical for the case with thermal light and was first discovered by Hanbury Brown and Twiss in 1956~\cite{brown1956correlation,brown1956test}.

In Ref.~\cite{zhang2019superbunching}, we proposed an effective way to introduce multiple indistinguishable two-photon paths in the spatial domain by establishing phase correlation among photons through a digital phase encoding technique. We encode the specially designed phase distribution dispersed randomly $N$ spatially correlated phases $\phi$, $2\phi$,~$\cdots$,~$N\phi$ onto the wave front of a coherent light, where $\phi$ is a temporally random phase with equal distribution probability within $[0, 2\pi)$. In this way, multiple different but indistinguishable two-photon paths can be introduced whenever the probability amplitudes of the two-photon paths $A_mD_1A_nD_2$ and $A_mD_2A_nD_1$ are of the same encoded random phase factor $(m+n)\phi=l\phi$ ($m, n=1,\,\cdots,\, N$), where $m\phi$ and $n\phi$ are the encoded random phase of photons $A_m$ and $A_n$, respectively. The probability amplitude of the two-photon path $A_mD_1A_nD_2$ can be expressed as $e^{im\phi}g_{m1}e^{in\phi}g_{n2}$. The coincidence counting rate can be calculated by summing over all probability contributions from all possible cases with different $l\phi$, and is given as
\begin{widetext}
\begin{eqnarray}\label{Eq:A2}
P^{(2)}_{(m,n)}&&=\left<\sum_{l=2}^{2N}\left|\sum_{m+n=l;m,n\in[1,N]}e^{i(m+n)\phi}g_{m1}g_{n2}\right|^2\right>\nonumber\\
&&=\left<\sum_{l=2}^{N+1}\left|\sum_{m=1}^{l-1}e^{im\phi}e^{i(l-m)\phi}g_{m1}g_{(l-m)2}\right|^2+\sum_{l=N+2}^{2N}\left|\sum_{m=l-N}^{N}e^{im\phi}e^{i(l-m)\phi}g_{m1}g_{(l-m)2}\right|^2\right>\, .
\end{eqnarray}
\end{widetext}

\noindent
Here in Eq.~(\ref{Eq:A2}), when there is only one term in $\left|\cdots\right|^2$, such as $g_{m1}g_{n2}$ with $m=n=1$,  it only contributes a constant background after taking the modulus. However, when there are two terms or more than two terms in $\left|\cdots\right|^2$,  such as $g_{m1}g_{n2}+g_{m2}g_{n1}+g_{p1}g_{q2}+g_{p2}g_{q1}$ satisfying $(m+n)\phi=(p+q)\phi=l\phi$, then the following three terms will appear after taking the modulus operation
\begin{widetext}
\begin{eqnarray}\label{Eq:A3}
e^{-\frac{ik}{z}x_1x'_m}e^{-\frac{ik}{z}x_2x'_n}e^{\frac{ik}{z}x_1x'_m}e^{\frac{ik}{z}x_2x'_n}=1,
\end{eqnarray}
\begin{eqnarray}\label{Eq:A4}
&&e^{-\frac{ik}{z}x_1x'_m}e^{-\frac{ik}{z}x_2x'_n}e^{\frac{ik}{z}x_1x'_n}e^{\frac{ik}{z}x_2x'_m}+e^{\frac{ik}{z}x_1x'_m}e^{\frac{ik}{z}x_2x'_n}e^{-\frac{ik}{z}x_1x'_n}e^{-\frac{ik}{z}x_2x'_m}\nonumber\\
&&=2\text{cos}\left(\frac{k}{z}(x_1-x_2)(x'_m-x'_n)\right),
\end{eqnarray}
\begin{eqnarray}\label{Eq:A5}
&&e^{-\frac{ik}{z}x_1x'_m}e^{-\frac{ik}{z}x_2x'_n}e^{\frac{ik}{z}x_1x'_p}e^{\frac{ik}{z}x_2x'_q}+e^{\frac{ik}{z}x_1x'_m}e^{\frac{ik}{z}x_2x'_n}e^{-\frac{ik}{z}x_1x'_p}e^{-\frac{ik}{z}x_2x'_q}\nonumber\\
&&=2\text{cos}\left(\frac{k}{z}\left(x_1(x'_m-x'_p)+x_2(x'_n-x'_q)\right)\right) \, .
\end{eqnarray}
\end{widetext}
\noindent
Here, $x_1$ and $x_2$ are the one-dimensional coordinates of the detectors $D_1$ and $D_2$, respectively. $x'_m$ ($m=1,\cdots,N$) is the one-dimensional coordinate of photons triggering a coincidence count. It should be mentioned that, as the wave front of the light is encoded with the spatially correlated phases, there is a constant factor $1/{\sqrt{N}}$ for each propagation factor. The contributions of the above three terms can be calculated by integrating respectively over the whole area of light source
\begin{widetext}
\begin{eqnarray}\label{Eq:A6}
P^{(2)}_{term1}(x_1,x_2)&&=\frac{1}{R^2}\iint\limits_{x'_m,x'_n\in R}\frac{1}{N^2}\frac{1}{z^4}\times1dx'_mdx'_n  \nonumber\\
&&=\frac{1}{N^2 z^4},
\end{eqnarray}

\begin{eqnarray}\label{Eq:A7}
P^{(2)}_{term2}(x_1,x_2)&&=\frac{1}{R^2}\iint \limits_{x'_m,x'_n\in R}\frac{1}{N^2}\frac{1}{z^4}\times2\text{cos}\left[\frac{k}{z}(x_1-x_2)(x'_m-x'_n)\right]dx'_mdx'_n\nonumber\\
&&=\frac{2}{N^2 z^4}\text{sinc}^2\left[\frac{kR}{2z}(x_1-x_2)\right],
\end{eqnarray}

\begin{eqnarray}\label{Eq:A8}
P^{(2)}_{term3}(x_1,x_2)&&=\frac{1}{R^4}\iint \limits_{x'_m,x'_n\in R,x'_p,x'_q\in R}\frac{1}{N^2}\frac{1}{z^4}\times 2\text{cos}\left(\frac{k}{z}\left(x_1(x'_m-x'_p)+x_2(x'_n-x'_q)\right)\right)dx'_mdx'_ndx'_pdx'_q \nonumber   \\
&&=\frac{2}{N^2 z^4}\text{sinc}^2\left(\frac{kx_1R}{2z}\right)\text{sinc}^2\left(\frac{kx_2R}{2z}\right).
\end{eqnarray}
\end{widetext}
\noindent
Note that the coefficient $1/R$  should be introduced to convert discrete summation into integration. Therefore, by substituting Eqs.~(\ref{Eq:A6}), (\ref{Eq:A7}) and (\ref{Eq:A8}) into Eq.~(\ref{Eq:A2}), the normalized second-order spatial correlation function can be written as
\begin{widetext}
\begin{eqnarray}\label{Eq:A9}
g_{N}^{(2)}(x_1,x_2)&&=\frac{\sum_{R}P^{(2)}_{(m,n)}}{P^{(1)}(x_1)P^{(1)}(x_2)}\nonumber\\
&&=\frac{1}{N^2}\sum_{l=2}^{N+1}\left\{(l-1)+2\left\lfloor\frac{l-1}{2}\right\rfloor\text{sinc}^2\left[\frac{kR}{2z}(x_1-x_2)\right]\right.\nonumber\\
&&\left.+\left[(l-1)(l-2)-2\left\lfloor\frac{l-1}{2}\right\rfloor\right]\text{sinc}^2\left[\frac{kx_1R}{2z}\right]\text{sinc}^2\left[\frac{kx_2R}{2z}\right]\right\}\nonumber\\
&&+\frac{1}{N^2}\sum_{l=N+2}^{2N}\left\{(2N-l+1)+2\left\lfloor\frac{2N-l+1}{2}\right\rfloor\text{sinc}^2\left[\frac{kR}{2z}(x_1-x_2)\right]\right.\nonumber\\
&&\left.+\left[(2N-l+1)(2N-l)-2\left\lfloor\frac{2N-l+1}{2}\right\rfloor\right]\text{sinc}^2\left[\frac{kx_1R}{2z}\right]\text{sinc}^2\left[\frac{kx_2R}{2z}\right]\right\}\nonumber\\
&&=\frac{1}{N^2}\sum_{l'=-N+2}^{N+1}\left\{|l'-1|+2\left\lfloor\left|\frac{l'-1}{2}\right|\right\rfloor\text{sinc}^2\left[\frac{kR}{2z}(x_1-x_2)\right]\right.\nonumber\\
&&\left.+\left[|l'-1||l'-2|-2\left\lfloor\left|\frac{l'-1}{2}\right|\right\rfloor\right]\text{sinc}^2\left[\frac{kx_1R}{2z}\right]\text{sinc}^2\left[\frac{kx_2R}{2z}\right]\right\},
\end{eqnarray}
\end{widetext}
\noindent
where $\left\lfloor\cdots\right\rfloor$ means to round the internal numbers downward.

When the light scattered by the phase plate 1 illuminates another phase plate 2 with similar phase distribution dispersed randomly $N'$ spatially correlated phases $\phi'$, $2\phi'$, $\cdots$, $N'\phi'$, where $\phi'$ is a temporally random phase with equal distribution probability within $[0, 2\pi)$, the wave front of light after the phase plate 2 will be encoded with an addition phase delay. If the two detectors $D_1$ and $D_2$ are placed behind the phase plate 2 for coincidence measurement, there are many indistinguishable two-photon paths $A_{mp}D_1B_{nq}D_2$ satisfying $m\phi+n\phi+p\phi'+q\phi'=$constant, and the probability amplitude of which can be expressed as
\begin{eqnarray}\label{Eq:A10}
&&F(A_{mp}D_1A_{nq}D_2)\nonumber\\
&&=e^{i(m\phi+p\phi')}g_{mp1}e^{i(n\phi+q\phi')}g_{nq2}\, ,
\end{eqnarray}
\noindent
which corresponds to the case that the photon $A_{mp}$ encoded with a random phase $m\phi+p\phi'$ by the phase plate 1($m\phi$) and the phase plate 2 $(p\phi')$ in sequence is detected by detector $D_1$ and the photon $A_{nq}$ encoded with a random phase $n\phi+q\phi'$ by the phase plate 1 ($n\phi$) and the phase plate 2 $(q\phi')$ in sequence is detected by detector $D_2$. Similarly, under the paraxial approximation, the total coincidence counting rate is the coherent superposition of all probability amplitudes and is written as
\begin{widetext}
\begin{eqnarray}\label{Eq:A11}
P_{(m,n,p,q)}^{(2)}&&=\left<\sum_{l=2}^{N+1}\sum_{l'=2}^{N+1}\left|\sum_{m=1}^{l-1}\sum_{p=1}^{l'-1}e^{i(m\phi+n\phi+p\phi'+q\phi')}g_{mp1}g_{nq2}\right|^2\right.\nonumber\\
&&+\sum_{l=2}^{N+1}\sum_{l'=N+2}^{2N}\left|\sum_{m=1}^{l-1}\sum_{p=l'-N}^{N}e^{i(m\phi+n\phi+p\phi'+q\phi')}g_{mp1}g_{nq2}\right|^2\nonumber\\
&&+\sum_{l=N+2}^{2N}\sum_{l'=2}^{N+1}\left|\sum_{m=l-N}^{N}\sum_{p=1}^{l'-1}e^{i(m\phi+n\phi+p\phi'+q\phi')}g_{mp1}g_{nq2}\right|^2\nonumber\\
&&\left.+\sum_{l=N+2}^{2N}\sum_{l'=N+2}^{2N}\left|\sum_{m=l-N}^{N}\sum_{p=l'-N}^{N}e^{i(m\phi+n\phi+p\phi'+q\phi')}g_{mp1}g_{nq2}\right|^2\right>.
\end{eqnarray}
\end{widetext}
\noindent
Here, we assume $N'=N$ for simplicity. Therefore, the normalized second-order spatial correlation function can be obtained by integrating over the whole areas of light sources, and one gets
\begin{widetext}
\begin{eqnarray}\label{Eq:A12}
g^{(2)}_{N,2}(x_1,x_2)&&=\frac{\sum_{R}P^{(2)}_{(m,n,p,q)}}{P^{(1)}(x_1)P^{(1)}(x_2)}\nonumber\\
&&=\frac{1}{N^4}\sum_{l=-N+2}^{N+1}\left\{|l-1|+2\left\lfloor\left|\frac{l-1}{2}\right|\right\rfloor\text{sinc}^2\left[\frac{kR}{2z}(x_1-x_2)\right]\right.\nonumber\\
&&\left.+\left[|l-1||l-2|-2\left\lfloor\left|\frac{l-1}{2}\right|\right\rfloor\right]\text{sinc}^2\left(\frac{kx_1R}{2z}\right)\text{sinc}^2\left(\frac{kx_2R}{2z}\right)\right\}\nonumber\\
&&\times\sum_{l'=-N+2}^{N+1}\left\{|l'-1|+2\left\lfloor\left|\frac{l'-1}{2}\right|\right\rfloor\text{sinc}^2\left[\frac{kR}{2z}(x_1-x_2)\right]\right.\nonumber\\
&&\left.+\left[|l'-1||l'-2|-2\left\lfloor\left|\frac{l'-1}{2}\right|\right\rfloor\right]\text{sinc}^2\left(\frac{kx_1R}{2z}\right)\text{sinc}^2\left(\frac{kx_2R}{2z}\right)\right\}.
\end{eqnarray}
\end{widetext}
\noindent
One sees that, by cascading two phase plates in sequence, $g^{(2)}_{N,2}(x_1,x_2)$ is the product of the normalized second-order spatial correlation functions induced by individual phase plate, and the two-photon superbunching effect is enhanced cumulatively.
If $M$ phase plates are cascaded, in which each phase plate is of independent phase distribution dispersed randomly $N$ spatially correlated phases, the normalized second-order spatial correlation function can be calculated as
\begin{widetext}
\begin{eqnarray}\label{Eq:13}
g^{(2)}_{N,M}(x_1, x_2)&&=\frac{1}{N^{2M}}\prod^{M}_{i=1}\sum_{l_i=-N+2}^{N+1}\left\{|l_i-1|+2\left\lfloor\left|\frac{l_i-1}{2}\right|\right\rfloor \mathrm{sinc}^2\left[\frac{kR}{2z}(x_1-x_2)\right]\right.\nonumber\\
&&\left.+\left[|l_i-1||l_i-2|-2\left\lfloor\left|\frac{l_i-1}{2}\right|\right\rfloor\right]\mathrm{sinc}^2\left(\frac{kx_1R}{2z}\right)\mathrm{sinc}^2\left(\frac{kx_2R}{2z}\right)\right\}.
\end{eqnarray}
\end{widetext}
\noindent

\begin{acknowledgments}
This work is financially supported by the National Key Research and Development Program of China (Grant No.~2022YFA1404604), the National Natural Science Foundation of China (NSFC) (Grant No.~12134007), the Higher Education Discipline Innovation Project (Grant No.~B23045), the Fundamental Research Funds for the Central Universities and the project ZR2024QA225 supported by Shandong Provincial Natural Science Foundation.
\end{acknowledgments}


%

\end{document}